\documentclass[conference]{IEEEtran}

\usepackage{graphicx}
\graphicspath{ {figures/} }
\usepackage{url}
\urldef\spamassassinurl\url{http://csmining.org/index.php/spam-assassin-datasets.html?file=tl_files/Project_Datasets/SpamAssassin%20data/ }
\usepackage[]{algorithm2e}
\usepackage{comment}
\usepackage{mathtools}
\usepackage{subcaption}
\usepackage{amsmath}
\captionsetup[subfigure]{width=0.85\textwidth}
\let\labelindent\relax
\usepackage{enumitem}
\usepackage{booktabs}
\usepackage{float}
\usepackage{stfloats}

\usepackage{fancyhdr}
\usepackage{hyperref}
\fancypagestyle{firstpage}{
  \fancyhf{}
  
  \fancyfoot[L]{{\em This paper was published in the Proceedings of the 2020 IEEE 14th International 
Conference on Semantic Computing (ICSC), pages 77-84, San Diego, CA, USA, 2020. \copyright 2020 IEEE\\
Link to article abstract in IEEE Xplore: https://doi.org/10.1109/ICSC.2020.00018}}
}
\pagestyle{plain}

\begin{document}

\pagenumbering{gobble}

\title{Early Forecasting of Text Classification Accuracy and F-Measure with Active Learning}

\IEEEoverridecommandlockouts

\author{\IEEEauthorblockN{Thomas Orth}
\IEEEauthorblockA{Department of Computer Science\\
The College of New Jersey\\
Ewing, NJ 08628\\
Email: ortht2@tcnj.edu}
\and
\IEEEauthorblockN{Michael Bloodgood}
\IEEEauthorblockA{Department of Computer Science\\
The College of New Jersey\\
Ewing, NJ 08628\\
Email: mbloodgood@tcnj.edu}}

\maketitle

\thispagestyle{firstpage}

\begin{abstract} \label{sec:abstract}
When creating text classification systems, one of the major bottlenecks is the annotation of training data. Active learning has been proposed to address this bottleneck using stopping methods to minimize the cost of data annotation. An important capability for improving the utility of stopping methods is to effectively forecast the performance of the text classification models. Forecasting can be done through the use of logarithmic models regressed on some portion of the data as learning is progressing. A critical unexplored question is what portion of the data is needed for accurate forecasting. There is a tension, where it is desirable to use less data so that the forecast can be made earlier, which is more useful, versus it being desirable to use more data, so that the forecast can be more accurate. We find that when using active learning it is even more important to generate forecasts earlier so as to make them more useful and not waste annotation effort. We investigate the difference in forecasting difficulty when using accuracy and F-measure as the text classification system performance metrics and we find that F-measure is more difficult to forecast.  We conduct experiments on seven text classification datasets in different semantic domains with different characteristics and with three different base machine learning algorithms. We find that forecasting is easiest for decision tree learning, moderate for Support Vector Machines, and most difficult for neural networks.  
\end{abstract}
\section{Introduction} \label{sec:introduction}

Text classification has been used in many different applications and is an important task in semantic computing \cite{mishler2017ICSC, hoi2006, janik2008, allahyari2014, kanakaraj2015}. Using machine learning yields text classification systems with high performance, however, the major bottleneck in constructing new text classification systems is the cost of producing the training data. There has been a great deal of interest in reducing the annotation bottleneck for constructing new text classification systems through the use of active learning \cite{lewis1994, bloodgood2009NAACL, beatty2019ICSC}. Active learning works by having the learner actively select the data that will be labeled with the goal of optimizing learner efficiency by requesting labeling effort where it is expected to be most useful \cite{hantke2017, bloodgood2010ACL, lee2012, mairesse2010, miura2016}. To realize the potential benefits of active learning, it is crucial to stop the learning process when additional labels will no longer be useful. Determining when to stop active learning is an area of active research \cite{bloodgood2009CoNLL, bloodgood2013CoNLL, schohn2000, zhu2008b, laws2008, vlachos2008, zhu2008a, altschuler2019}. 

A related area of interest is to devise methods that can predict, or forecast, the performance of a machine learning model during learning. Accurate performance forecasting can improve our ability to determine when to stop seeking additional labeled data during active learning. Model forecasting can be done by performing regression on the performance of a machine learning model as more data is given to it. Prior work has shown the learning curve of a machine learning model has a shape similar to that of certain families of equations \cite{frey1999, singh2005}.

\begin{figure}
\centering
\includegraphics[width=0.50\textwidth]{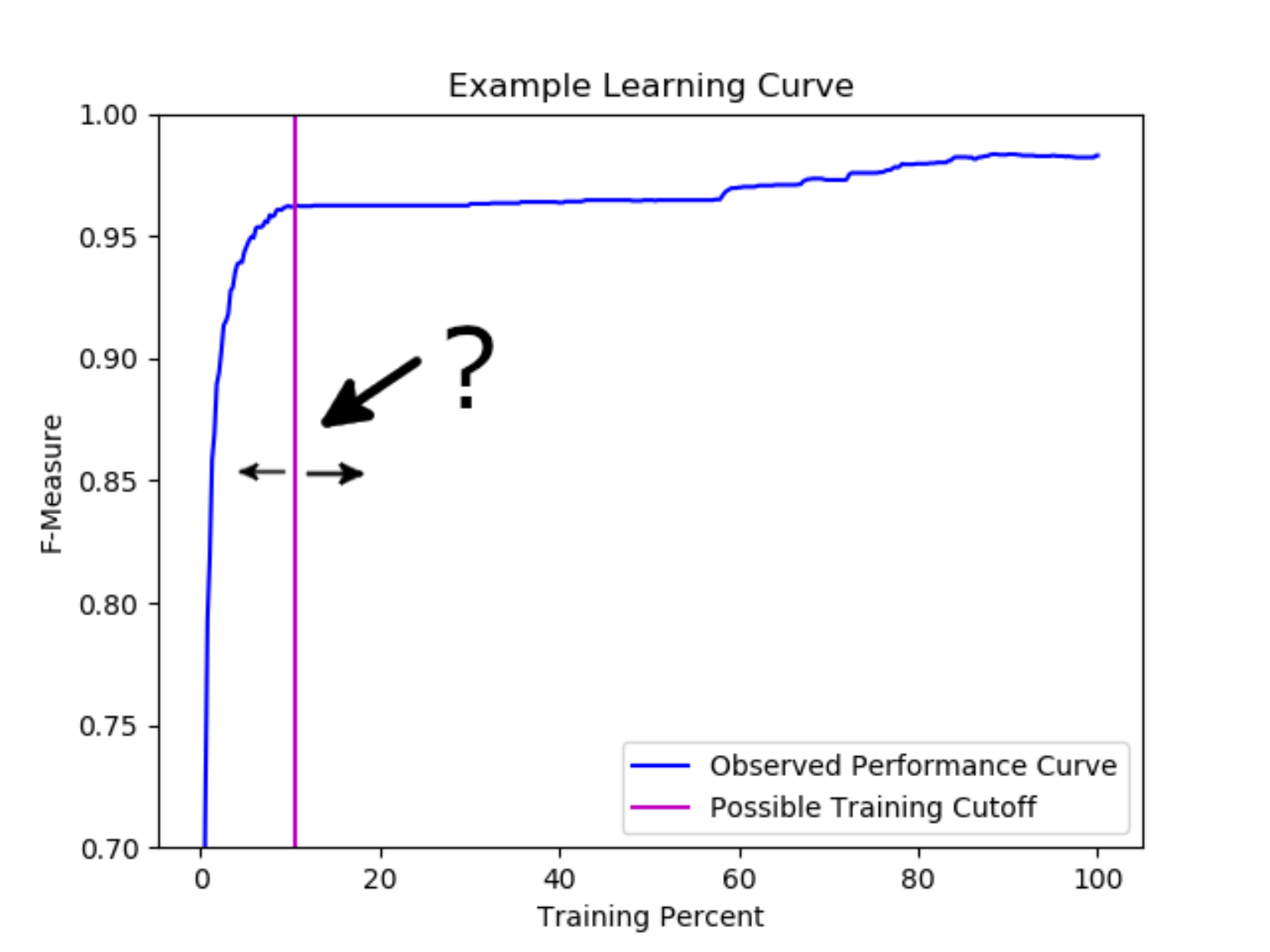}
\caption{Example Learning Curve showing the uncertainty of what value to use for the $TPC$ } \label{fig:start}
\end{figure}

\begin{figure}
\centering
\includegraphics[width=0.50\textwidth]{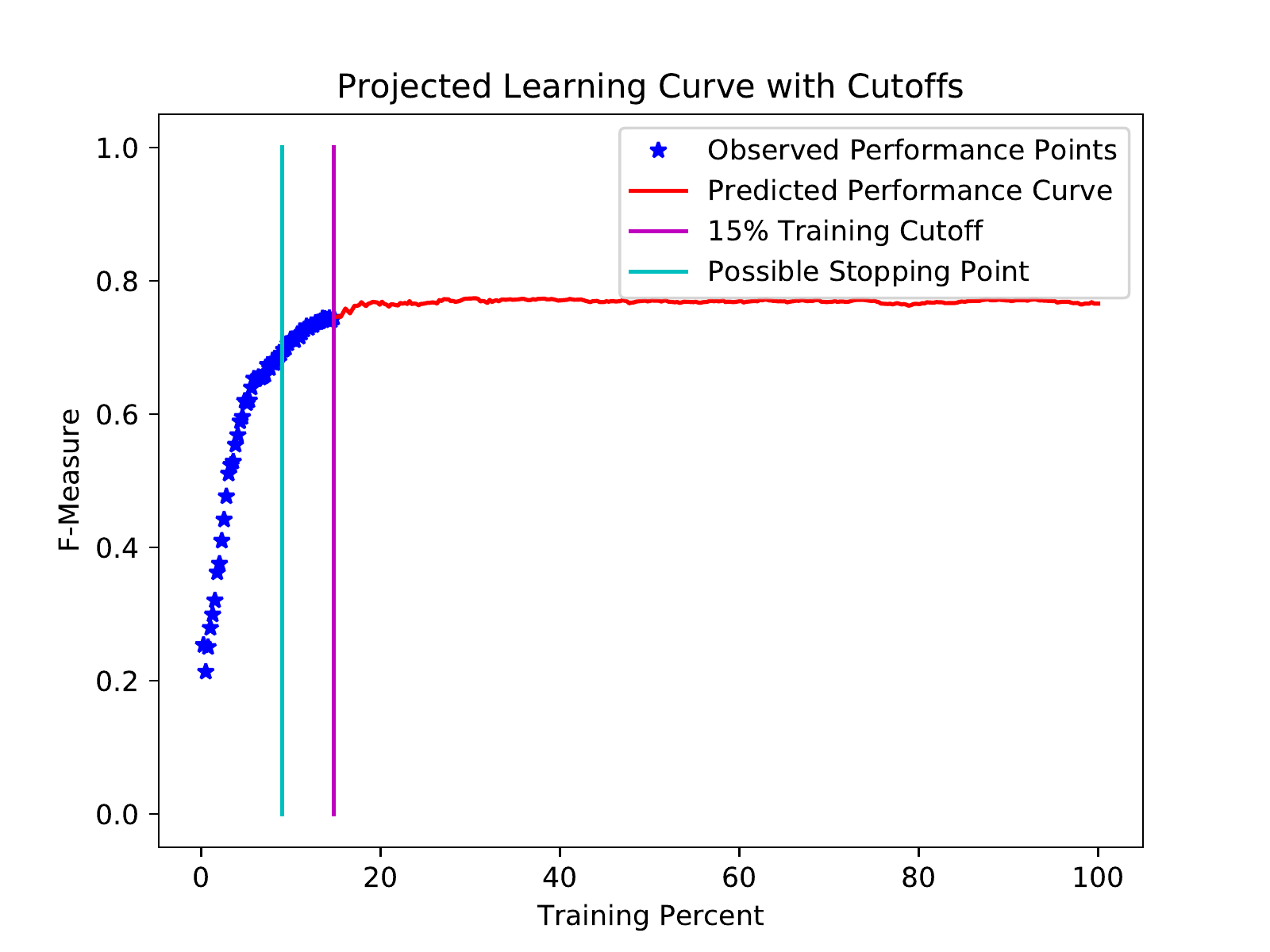}
\caption{Example curve demonstrating the prediction of performance using regression} \label{fig:proj}
\end{figure}

Figure \ref{fig:start} shows an example learning curve. Some part of the data is needed to create the forecasting model. The amount of points used to create the forecaster is determined by a Training Percent Cutoff ($TPC$). However, it is an open question where a good $TPC$ would be. In past work, 15\% has emerged as a pseudo-standard for setting the value of $TPC$ \cite{frey1999, singh2005}. However, setting the $TPC$ at 15\% might gather more labeled data than is necessary, wasting annotation effort. Figure~\ref{fig:proj} illustrates this with a hypothetical stopping point, shown by the leftmost vertical line in the figure. In this case the cutoff of 15\% would be wasteful of annotations because we would want to have stopped learning before we even are able to create the forecast that is supposed to help us determine when to stop learning. Section \ref{sec:results} shows actual stopping points for text classification, using a state of the art stopping method for active learning, are often well before 15\% of the data has been annotated.

In this paper, we explore the impact of different values for the $TPC$ to see how early we can forecast performance without losing a lot of accuracy in our forecasting. We also compare the $TPC$ value to the stopping percent found by a leading state of the art stopping method for active learning. We found that a smaller $TPC$ could be used for developing forecasting models than the 15\% that is currently widely used.  While this earlier forecasting is an improvement, we find that forecasting models are still not effective until too late compared to the stopping percent determined by stopping methods. This indicates that further research to make even earlier forecasting more accurate would be productive. We investigate forecasting model performance in terms of accuracy and F-Measure in section \ref{sec:acc-fm} and find that F-measure is significantly harder to forecast than accuracy. We explore the impact of using different batch percents as compared to using just 1\% and find that it makes little difference to forecasting capabilities. Additionally, we compare different base machine learning models and find that neural networks are more difficult to forecast than decision trees and SVMs (Support Vector Machines) are of medium difficulty to forecast. Finally, we compare passive and active learning and find that it is harder to forecast performance when using active learning.

\section{Related Work} \label{sec:relatedWork}

There has been a lot previous work in the area of using forecasting models to forecast the performance of base learners. Of these models, linear, exponential, logarithmic and power were the most popular.  Another model was proposed by Weiss and Tian, which has no name so hereafter will be referred to as the Weiss and Tian model \cite{weiss2008}.

There has been some research into creating systems that use specific parameters related to a base machine learning model in order to forecast performance. Past work has explored using hyperparameters of bayesian neural networks to forecast accuracy \cite{domhan2015, klein2017}.  This methodology does not work for our setting.  We forecast the performance of machine learning models based off the training data.

Other systems forecast the performance of machine learning models using task specific information. For the task of machine translation, past work has investigated forecasting performance by using a feature vector of information such as average sentence length of the test set \cite{kolachina2012}.  These methods utilize a lot of properties that are specific to machine translation, which cannot easily be adapted to other NLP tasks such as text classification.

There have been systems that use training set percentage to forecast machine learning model performance.  This was first done by Frey and Fisher in 1999 by forecasting decision trees \cite{frey1999}. Frey and Fisher used 15\% of the points of a learning curve to train their forecasting models and tested the model on the other 85\%.  They came to the conclusion that power law was the best way to forecast a learning curve.  We explore whether the $TPC$ can be varied and find that in practical active learning situations a $TPC$ less than 15\% would be desired. 

Singh investigated forecasting further and used different machine learning models in his experiments and found that logarithmic models worked better for forecasting than the power models \cite{singh2005}. Hence, we focus on logarithmic models in our paper. Finally, there has been some work from other areas such as using projective sampling to reduce the total cost of data mining that looks at using different amounts of training points for forecasting machine learning performance \cite{last2009}. We investigate this as well, but with a more fine-grained examination of how the $TPC$ can be varied for forecasting text classification performance as measured by different performance metrics with both active learning and passive learning. 

There has been some prior work in predicting performance in active learning.  Figueroa et al. performed a comparison between passive and active learning for predicting performance \cite{figueroa2012}.  In our experiments, we not only compare passive and active learning, but we also examine how much data to use for the forecasting process.

There has been some prior work in stopping active learning using mathematical guidance on how much performance can be expected to change.  For example, some past work has investigated how mathematical bounds on the amount of possible change in F-Measure from iteration to iteration can be used to stop the training process during active learning \cite{bloodgood2013CoNLL, altschuler2019}. Their method doesn't require labeled data to measure performance. In contrast, our method in the current paper requires some labeled data to obtain the initial points we use to regress our models. However, our models can then forecast levels of performance in addition to only changes. Future work includes developing algorithms to combine the mathematical bounds approach of \cite{bloodgood2013CoNLL, altschuler2019} with the regression approach in the current paper.  
\section{Experimental Setup} \label{sec:experimentalSetup}

We now will describe our experimental setup. All of our experiments are conducted in an iterative learning setting, which we describe in section~\ref{sec:iter}. 

\subsection{Iterative Learning Setup} \label{sec:iter}
We use the 20NewsGroups dataset\footnote{Downloaded the ``bydate" version from \url{http://qwone.com/~jason/20Newsgroups/}.  This version does not include duplicate posts and is sorted by date into train and test sets.}, the Reuters dataset, in particular the Reuters-21578 Distribution 1.0 ModApte split\footnote{ \url{http://www.daviddlewis.com/resources/testcollections/reuters21578/}} as done in \cite{joachims1998} and \cite{dumais1998}, the WebKB dataset\cite{mccallum1998}, the spamassassin corpus\cite{sculley2007}, the IMDB sentiment dataset\footnote{\url{http://ai.stanford.edu/~amaas/data/sentiment}}, TrecSpam 2005 ham25\cite{cormack2005}, and the first 20000 entries of Ohsumed\footnote{Downloaded from http://disi.unitn.it/moschitti/corpora.htm on July 13, 2017} for our experiments. We report the results for the four largest categories of the WebKB
dataset as done in past work \cite{mccallum1998, zhu2008a, zhu2008b}. We used 10-fold cross validation and present the averages for SpamAssassin, WebKB, Trec and Ohsumed.  For the other datasets, we used the standard train-test split provided by the dataset. For text classification, we use a bag of words approach with a frequency cutoff of three, meaning that each feature is a word and we don't create features for words that occur fewer than three times. We use binary feature values, meaning the value of the feature is a 1 if the feature (word) occurs in the document and 0 if the feature (word) does not occur in the document. There are also words that hold very little to no value for classification, called stop words.  We remove stop words that appear in the \textit{Long Stopword List} from https://www.ranks.nl/stopwords. We use SVM, decision tree, and multi-layer perceptron neural network as our main classifiers.  For SVM, we use a linear kernel.  For our neural network, we use a densely connected layer for the input layer with 64 hidden units, a dropout layer with 20\% dropout and another densely connected layer for the output with a hidden unit.  We used a passive selection algorithm which chooses random samples with all three base learners. We also use the closest-to-hyperplane selection algorithm with SVM for active learning \cite{campbell2000, tong2001, schohn2000}. This is because previous work has shown that it has better performance over other selection algorithms used \cite{bloodgood2018ICSC}.  For each iteration of training, the number of samples used is determined by a batch percent, $bp$. The $bp$ percent of the total amount of unlabeled data originally available is the amount of data that will be added to the labeled training data during each iteration of learning. We use different batch percents in our experiments. We first used 1.0\% to compare to Frey and Fischer \cite{frey1999}. We also used 0.25\% to test whether more fine-grain sampling would have any impact on forecasting performance.

\subsection{Overview of Predicting Learning Curves}

\begin{figure}
\centering
\includegraphics[width=0.50\textwidth]{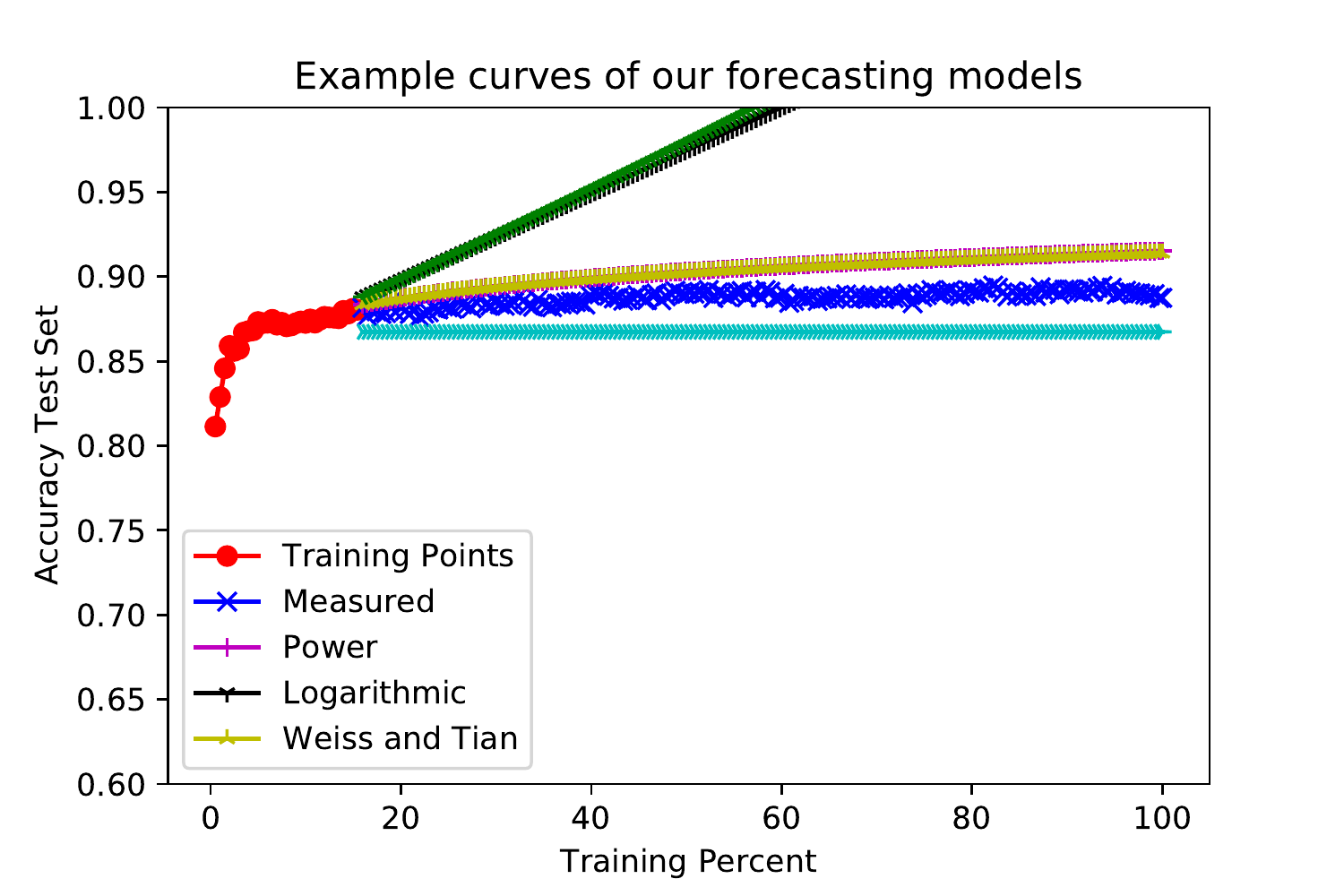}
\caption{Example curves of the different forecasting models} \label{fig:curves}
\end{figure}

Figure \ref{fig:curves} shows an example, using the described setup to forecast the performance of a decision tree on our Ohsumed dataset, where in this case our performance metric is accuracy.  Linear, Weiss and Tian, and exponential were the least accurate when used to forecast performance.   Power and logarithmic models do the best. The equations are shown in Table \ref{tab:equations}. In past literature, logarithmic was found to be the best forecaster \cite{singh2005}. In our experiments, we also observed logarithmic to be the best forecasting model. Therefore, for all the rest of the experiments in this paper, we use logarithmic as our forecasting model.

\subsection{Forecasting Performance Setup} \label{sec:pred}
\begin{table}
\centering
\begin{tabular}{lr}
Name & Equation\\
\hline
Linear & $y = ax + b$\\
Power & $y = a * x^b$\\
Logarithmic & $y = alog(x) + b$\\
Exponential & $y = a10^{bx}$\\
Weiss and Tian & $y = a + bx/(x+1)$\\
\end{tabular}
\caption{List of equations used}
\label{tab:equations}
\end{table}

Once we run the iterative learning process described in Section \ref{sec:iter} and record the performance of the learned classifier on held-out test data at each iteration, we then process all of the data to forecast the performance of the machine learning models used. We use the equations specified in Table \ref{tab:equations} to forecast on the given data. Notice in the equations the variables $y$, $x$, $a$, and $b$.  The first variable, $y$, is our classification performance metric. We experimented with two performance metrics, Accuracy and F-Measure\footnote{Both Accuracy and F-Measure are commonly used performance metrics for evaluating text classification performance. Accuracy is the percentage of classifications that are correct, while F-Measure is the harmonic mean of Precision and Recall, with Precision defined as the percentage of predicted positive instances that are truly positive instances and Recall defined as the percentage of truly positive instances that are predicted as positive instances.} The next variable, $x$, is the parameter from the data we collected that we are using to perform our prediction. For our experiments, we used the training percent for the current iteration, consistent with past work \cite{frey1999, singh2005}. The final two parameters, $a$ and $b$, are the learned coefficients from performing regression on the given data.  

In order to evaluate the performance of forecasting models, we define a measurement, which we call Average Difference, that captures how much the forecasted values differ on average from the observed values. Average Difference is defined in equation~\ref{equ:avg_diff} below. 
\begin{equation}
\textrm{Average Difference} = \frac{\sum\limits_{i=1}^{n} |f(x_{i}) - y_i |}{n}
\label{equ:avg_diff}
\end{equation}
where $f$ is the forecasting function, $x_{i}$ is the training percent at the $i^{th}$ test point, $y_{i}$ is the observed performance at the $i^{th}$ point, and $n$ is the number of test points, as defined in equation~\ref{equ:n} below.
\begin{equation}
n = \frac{100-TPC}{bp}
\label{equ:n}
\end{equation}
where $TPC$ is the Training Percent Cutoff and $bp$ is the batch percent, both as defined earlier in this paper.

\begin{figure}
\centering
\includegraphics[width=0.50\textwidth]{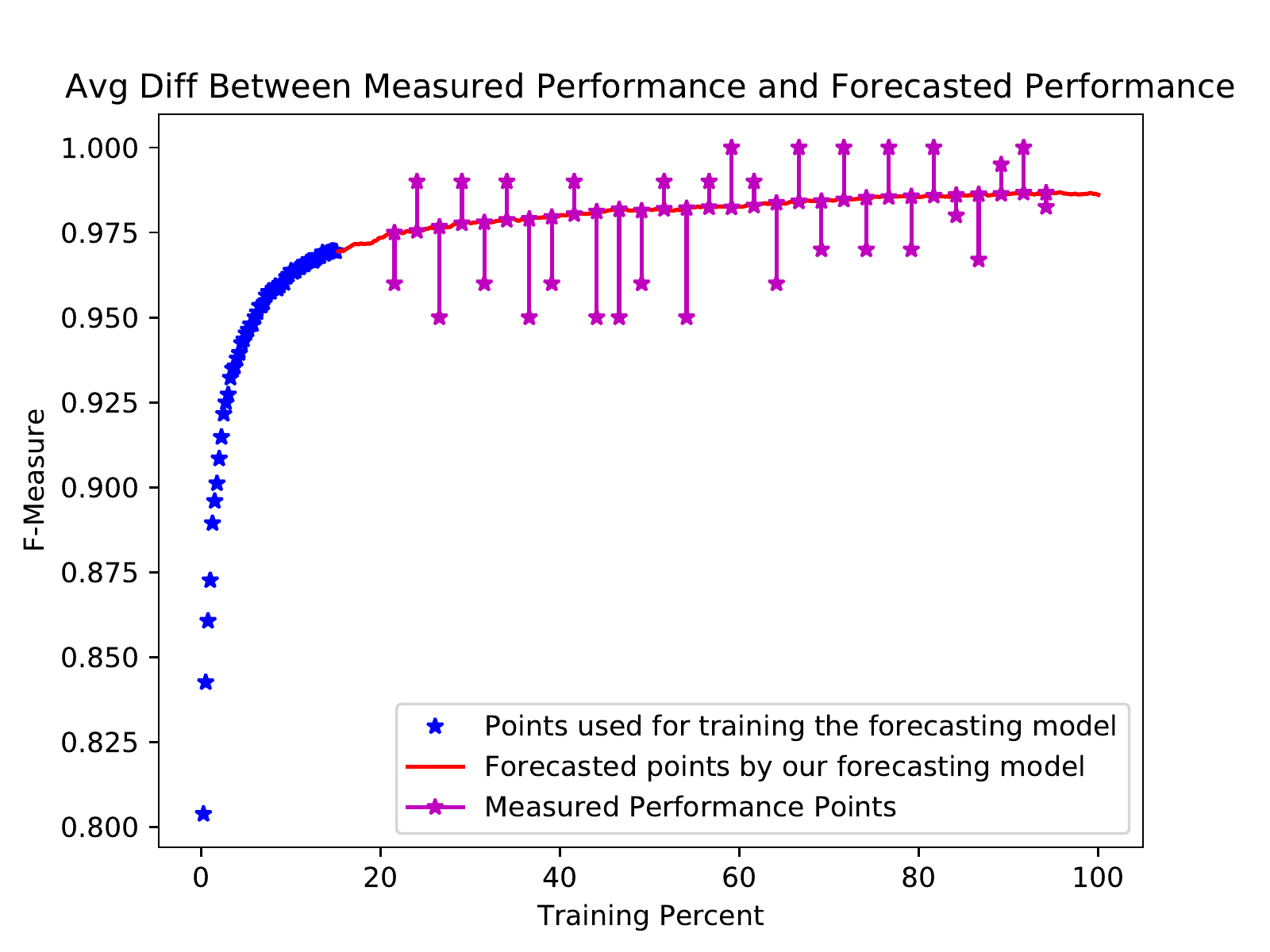}
\caption{Illustration of Average Difference} \label{fig:avg_difference}
\end{figure}

Average Difference is illustrated in Figure \ref{fig:avg_difference}. The points lying around our predicted curve are example measured points.  Each point is of the form $(x_i,y_i)$. We use equation~\ref{equ:avg_diff}. We go over all points from $i=1$ to $n$, with $x_1$ being the x-coordinate of the first point after the $TPC$.  We take the difference of the forecasted performance and the observed performance at each point and average these differences.  When the Average Difference measurement is smaller, that means our forecasting model is performing better.
\section{Results} \label{sec:results}

This section discusses the results of our experiments. We show the impact of batch percent on forecasting performance, we compare the performance of forecasting classification performance in terms of Accuracy versus in terms of F-Measure, we analyze varying the $TPC$, we show how the choice of base learner impacts forecasting, and finally we show the difference between forecasting in a passive learning setting versus in an active learning setting.

\subsection{Impact of Batch Percent}

We performed a comparison of forecasting performance when batch percent is 0.25\% versus when batch percent is 1\%, by computing the Average Difference as defined in equation \ref{equ:avg_diff}. We compute an overall average difference by averaging the individual average differences over all datasets and base machine learning models.  We use 15\% as our $TPC$ as it is a commonly used $TPC$ \cite{frey1999, singh2005}. For Accuracy, the overall average difference was 0.0256 for 0.25\% batch percent and 0.0208 for 1.0\%.  We can see that there is not much difference between the change in batch percent for accuracy. For F-Measure, the overall average difference was 0.167 for 0.25\% batch percent and 0.129 for 1.0\% batch percent. Again, we see there is not much difference in performance of our forecasting system when we vary batch percent. It is possible that with much larger batch percents we would see a change in forecasting performance, but using larger batch percents is known to have various negative effects on active learning \cite{beatty2018ICSC, brinker2003}, so we did not investigate the impact with larger batch percents that are less likely to be used in practice. 

\subsection{Accuracy vs. F-Measure} \label{sec:acc-fm}

\begin{figure}
\centering
\includegraphics[width=0.5\textwidth]{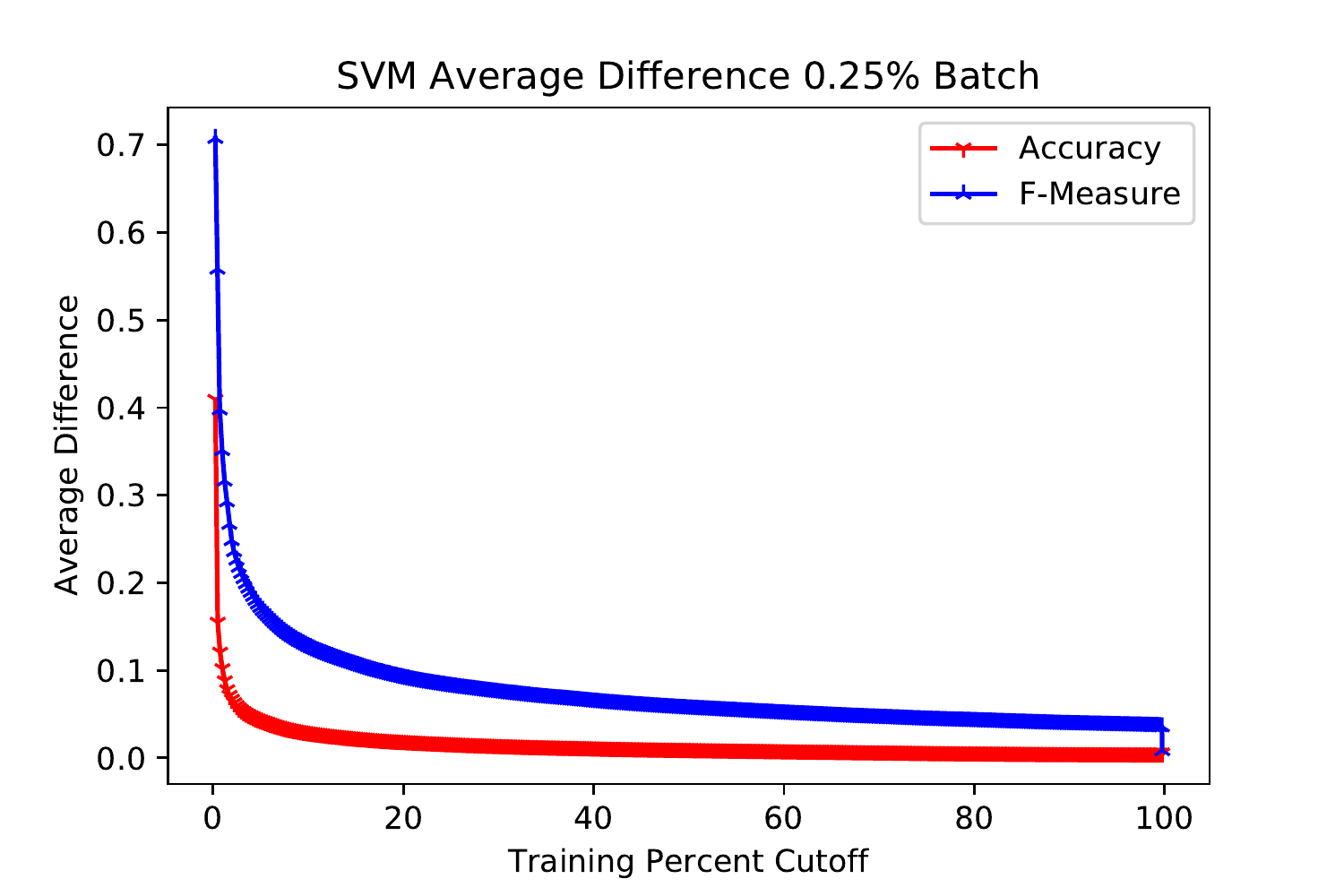}
\caption{Quality of forecasting (as measured by Average Difference) for varying $TPC$ values when text classification performance is measured in terms of Accuracy and in terms of F-Measure. A lower Average Difference means a higher quality forecast.} \label{fig:acc_fm}
\end{figure}

In this section we compare forecasting when text classification performance is measured in terms of Accuracy versus in terms of F-Measure. In these experiments we vary the $TPC$ to go through all possible $TPC$ values. Figure~\ref{fig:acc_fm} shows the overall average difference for Accuracy and F-Measure using SVM as the base learner over all datasets using 0.25\% as the batch percent. The results are compelling: Accuracy has a much lower average difference than F-Measure. This shows that new forecasting methods are needed when classification performance is measured in terms of F-Measure.

\subsection{$TPC$ Analysis} \label{sec:train}

Informally it is expected that forecasting is more useful if it can be done earlier in the iterative training process, however, it is also expected to be more difficult to create high quality forecasts earlier in the process. In this section, we examine these issues in detail, with experiments illuminating more specifically the value of forecasting by certain points in the iterative learning process and the expected changes of the quality of the forecasts due to changes in when the forecast is created. Specifically, we experiment with changing the $TPC$. Frey and Fisher used 15\% for the $TPC$ \cite{frey1999}. There was no analysis done of possibly changing the $TPC$. We use a 0.25\% batch percent for the experiments in this section. 

Figure~\ref{fig:acc_fm} shows the Average Difference using SVM as the base learner and varying the $TPC$ from 0.25\% of our data to 99.75\% of our data to show a fine-grained look at the impact of changing the $TPC$ on forecast quality when Accuracy is used as the classification performance metric and when F-Measure is used as the classification performance metric. Figure~\ref{fig:acc_fm} shows that as $TPC$ is increased, our forecasting quality improves, or in other words, our average difference gets smaller. However, we can see that the rate of improvement in forecasting quality is very different at different points in the iterative learning process, or in other words, at different $TPC$ values. In particular, there is a very sharp improvement in forecasting quality up to about ten percent $TPC$ and then the rate of improvement is much slower, with forecasts improving only by small amounts for larger settings of the $TPC$. This shows that the $TPC$ can potentially be pushed back a bit lower than 15\% without sacrificing too much forecast quality, especially for Accuracy. For F-Measure, the shape is not as much of an elbow dip, but as discussed in section~\ref{sec:acc-fm}, current forecasting methods don't work well for F-Measure and are in need of improvement. 

\begin{figure}
\centering
\includegraphics[width=0.5\textwidth]{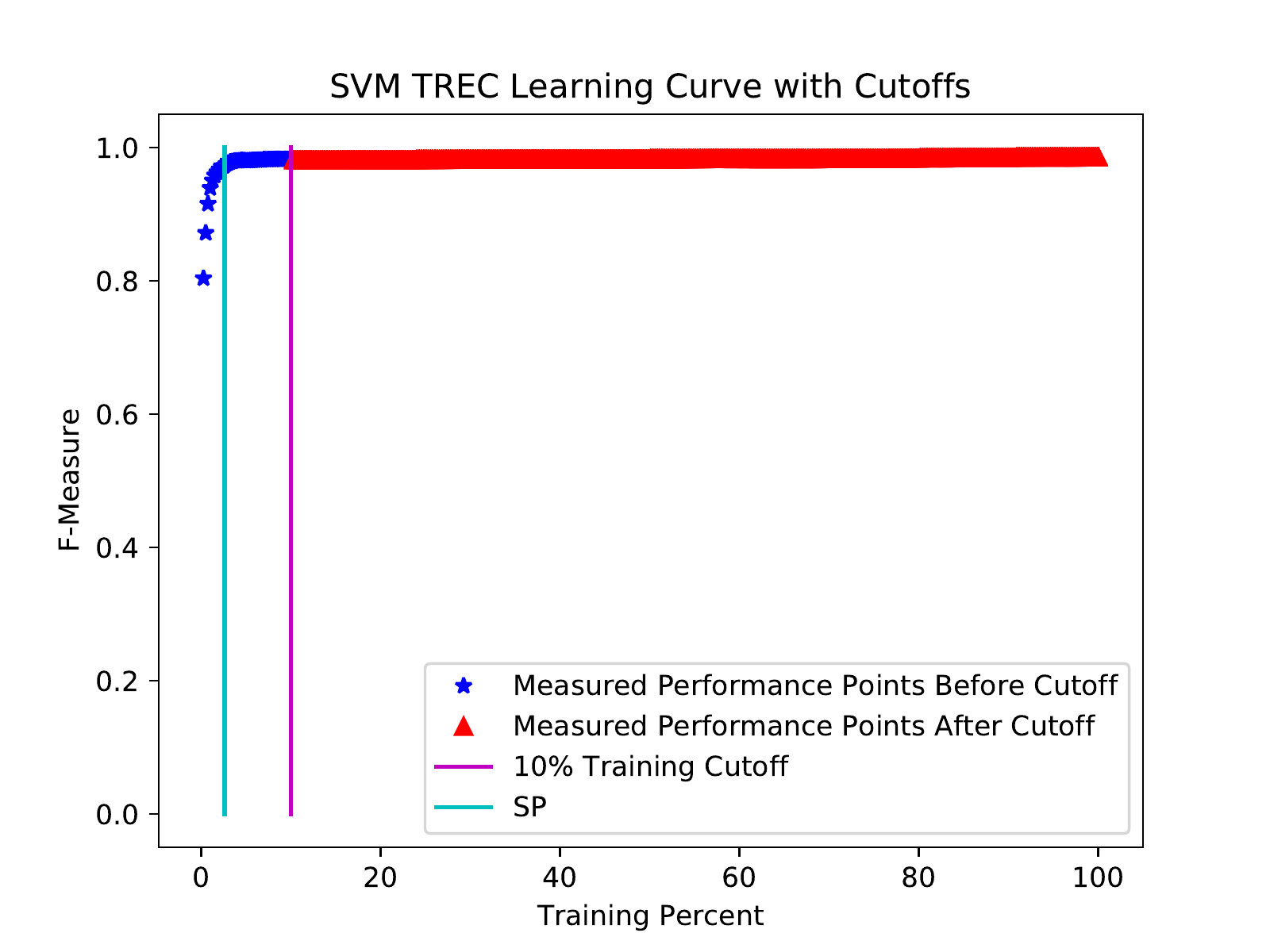}
\caption{Learning Curve using active learning with SVM and closest-to-hyperplane sampling on the TREC dataset. The $TPC$ is set to 10\%, about the earliest the current state-of-the-art can be set to without sacrificing large amounts of forecast quality. The stopping percent automatically determined during active learning by the Stabilizing Predictions (SP) method from \cite{bloodgood2009CoNLL} is shown by the SP vertical line.} \label{fig:new_cut}
\end{figure}

Table~\ref{tab:stop} shows the stopping points automatically determined during active learning for all of our datasets. These results were obtained by using the state-of-the-art stopping method for active learning described in \cite{bloodgood2009CoNLL}, hereafter referred to as the Stabilizing Predictions (SP) method. The stopping point percents in Table~\ref{tab:stop} were determined using an active learning (or in other words, selective sampling) setting with SVM as the base learner and closest-to-hyperplane sampling as the selection algorithm. Figure~\ref{fig:new_cut} shows the situation for the TREC dataset. In Figure~\ref{fig:new_cut} the $TPC$ is reduced to 10\% from the previously used 15\% since our results showed it could be pushed back to about 10\% without sacrificing large amounts of forecast quality. However, the stopping percent is even smaller than this reduced $TPC$, showing that it would be practically valuable to develop new forecasting methods that can forecast with higher quality earlier than the current state-of-the-art approach. 

\begin{table}[htb]
    \centering
    \begin{tabular}{@{}cccc@{}}
        \toprule
        Dataset & Stopping Percent \\
        \midrule
        20NewsGroups & 5.922 \\
        Reuters & 4.795 \\
        Ohsumed & 11.824 \\
        SpamAssassin & 5.109 \\
        Trec & 2.605 \\
        WebKB & 11.765 \\
        IMDB & 15.996 \\
    \end{tabular}
    \caption{Stopping Percents automatically determined during active learning by using the Stabilizing Predictions (SP) method from \cite{bloodgood2009CoNLL} \label{tab:stop}}
\end{table}

\subsection{Impact of Base Learner on Forecasting Performance}

\begin{figure}
\centering
\includegraphics[width=0.5\textwidth]{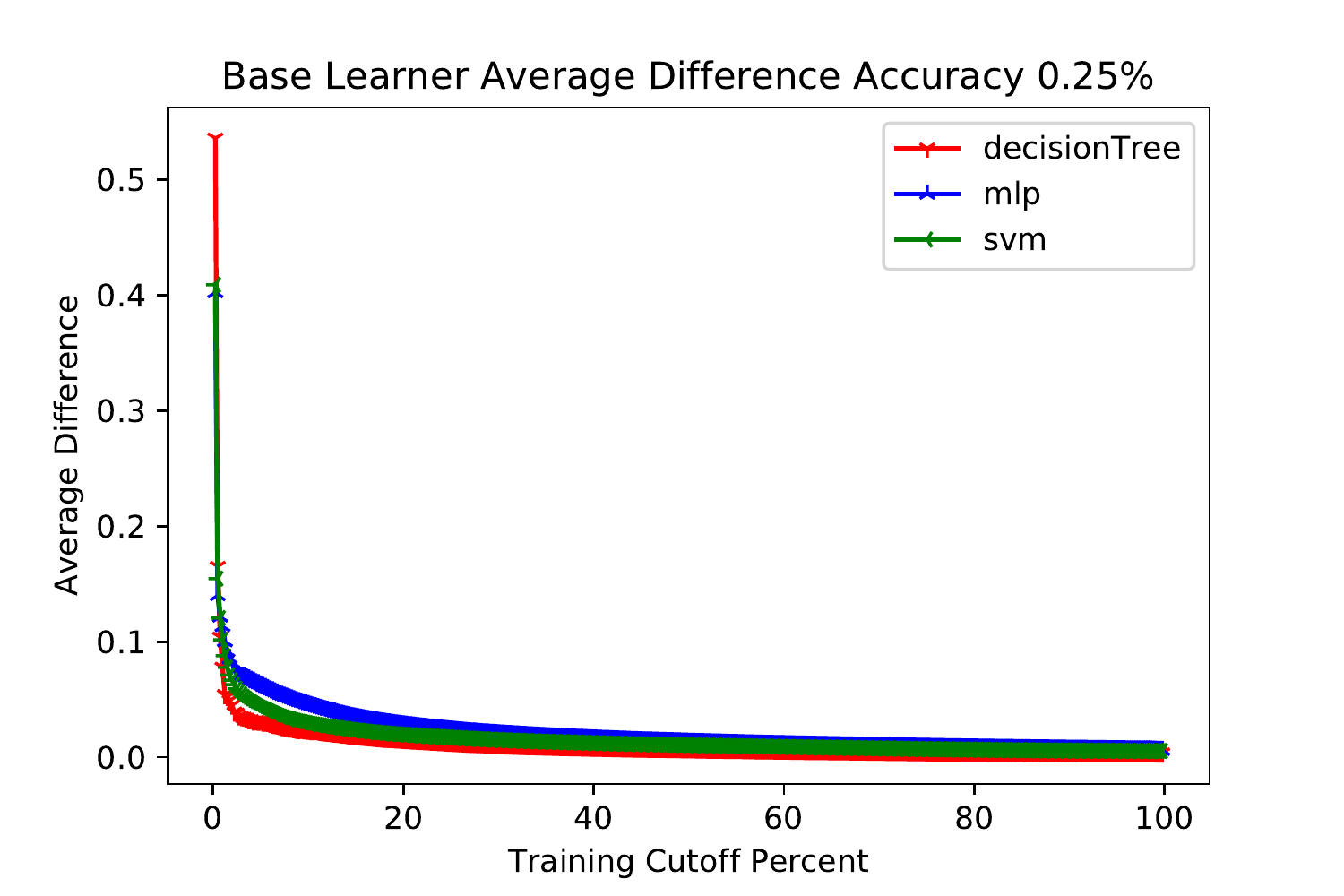}
\caption{Overall Average Difference over all datasets when classification performance is measured by Accuracy for different base learners. Decision tree classifiers are the easiest to forecast, neural network classifiers are the hardest to forecast, and SVM classifiers are in the middle.} \label{fig:glob_model}
\end{figure}

This section shows the impact of the base learner (SVM, decision tree, neural network) used during iterative learning.  Results are only shown for Accuracy as F-Measure curves represented similar results. Figure~\ref{fig:glob_model} shows the overall average difference of the forecasts for the different base machine learning models for varying $TPC$ values.  As shown, decision tree classifiers are the easiest to forecast, neural network classifiers are the hardest to forecast, and SVM classifiers are in the middle. 

\subsection{Impact of Passive Learning vs Active Learning}

The results in this section show how well forecasting can be done in a passive learning setting versus in an active learning setting. For passive learning, we randomly select the next batch of data to be labeled at each iteration of the iterative learning process described in section~\ref{sec:iter}. This is the standard setting under which most forecasting methods have been developed and tested \cite{singh2005, frey1999}. 

For active learning, an algorithm actively selects the next batch of data it wants to have labeled at each iteration of the iterative learning process. The idea is that by selectively sampling the examples the algorithm expects to be most valuable to have labeled, an effective model will be able to be learned from smaller amounts of data, thereby reducing data labeling cost. Since forecasting is intended to be used to help provide guidance on when to stop labeling additional data so that data labeling efforts are not wasted, it is a natural fit that forecasting could be of particular value and interest in active learning settings. However, investigations of forecasting effectiveness in active learning settings have been limited. 

We have already seen that SVM is the middle base learner in terms of forecasting difficulty. Furthermore, active learning has been well studied with SVMs and a well known successful algorithm is to sample the examples that are closest to the current model's learned hyperplane as was discussed in section~\ref{sec:iter}. For these reasons the results we present in this section are for SVM with passive learning versus for SVM with active learning as implemented by the closest-to-the-hyperplane selection algorithm. Also, all results in this section are for forecasting classification performance in terms of Accuracy since forecasting performance in terms of F-Measure is an area in need of future work. 

\begin{figure}
\centering
\includegraphics[width=0.5\textwidth]{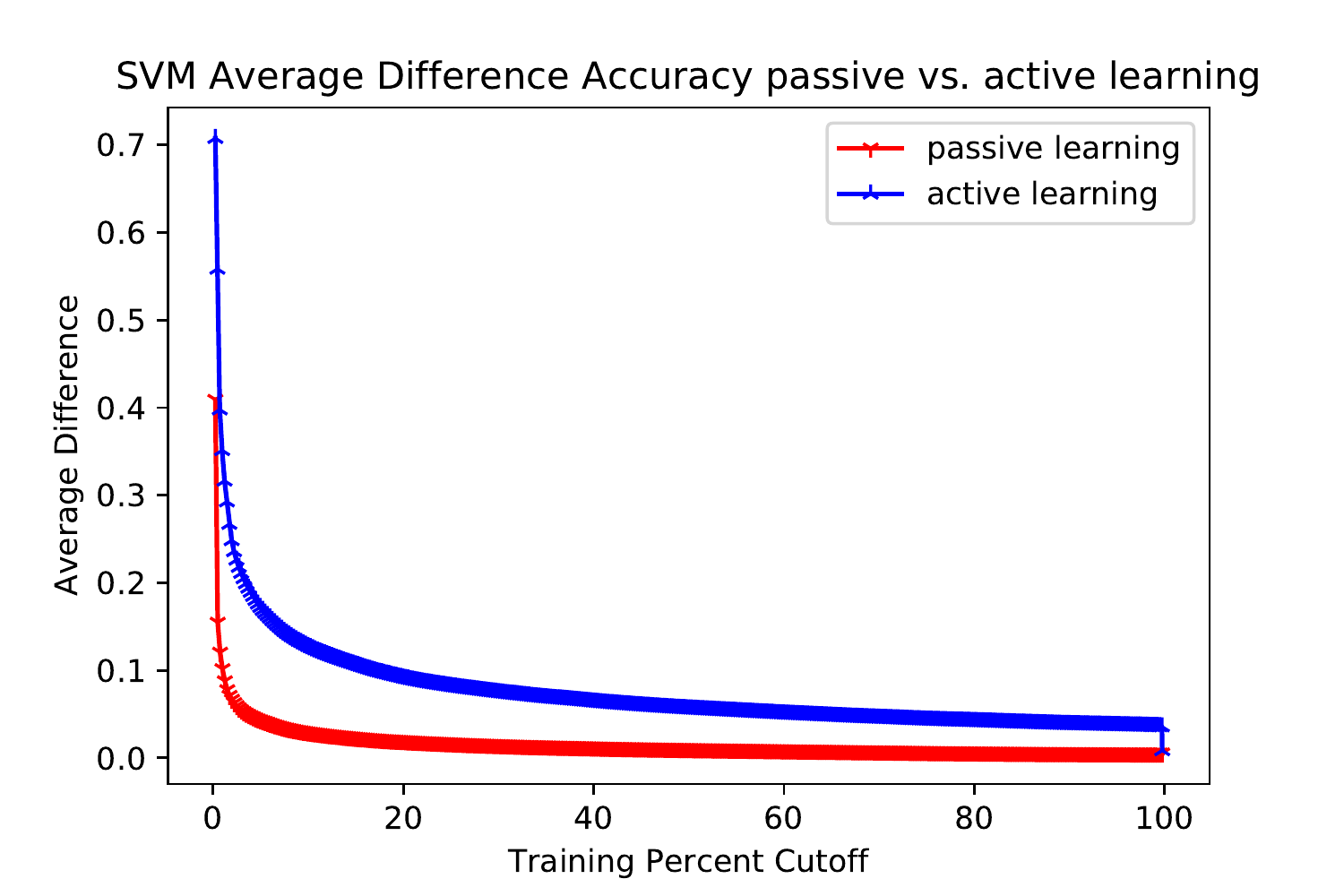}
\caption{Overall Average Difference over all datasets when classification performance is measured by Accuracy for SVM base learner in a passive learning setting (random selection of examples at each iteration) and an active learning setting (closest-to-hyperplane selection of examples at each iteration). The results show that current forecasting methods work much better in a passive learning setting. Lower Average Difference means higher quality forecast.} \label{fig:sampling}
\end{figure}

\begin{figure}
\centering
\includegraphics[width=0.5\textwidth]{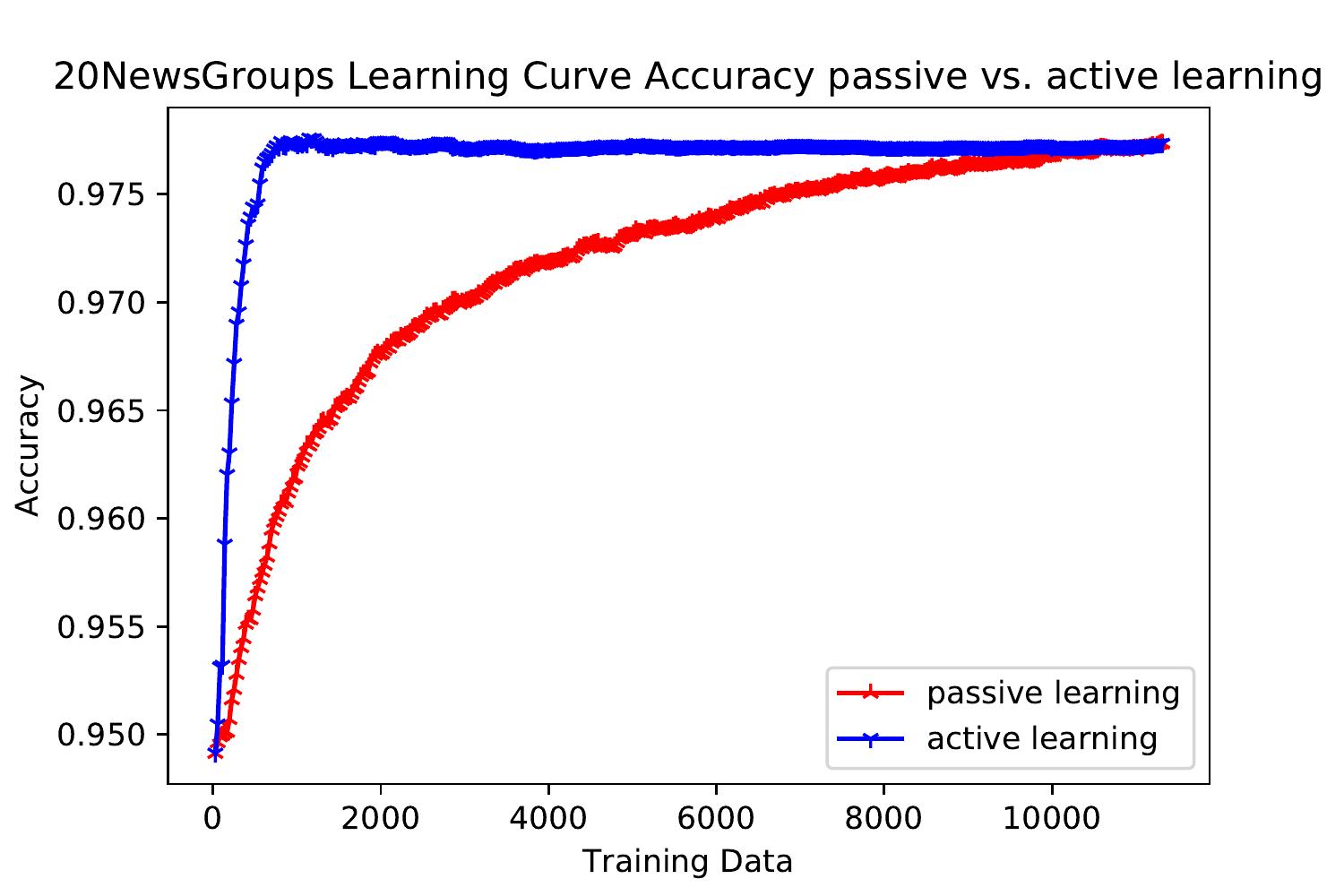}
\caption{Learning curves for 20NewsGroups dataset when classification performance is measured by Accuracy for SVM base learner for passive learning (random selection of examples at each iteration) and active learning (closest-to-hyperplane selection of examples at each iteration). The active learning curve deviates from a logarithmic shape making it difficult for existing state-of-the-art forecasting methods to generate high quality forecasts in active learning settings.} \label{fig:curve_sampling}
\end{figure}

Figure~\ref{fig:sampling} shows compelling results: current state-of-the-art forecasting methods perform much better when using passive learning than when using active learning. To see why this is the case, we show the learning curves for each setting. Figure \ref{fig:curve_sampling} shows the learning curves for the 20NewsGroups dataset when using passive learning and active learning. The active learning curve has a different shape, deviating from the shape of a logarithmic curve.  Because of this, it's harder to forecast the performance of SVM with active learning by assuming that a learning curve shape follows the shape of a logarithmic function. This shows the need for improving the state-of-the-art so that we can forecast effectively in active learning settings. Future work that could be promising for accomplishing this includes developing algorithms to combine the mathematical bounds approach of \cite{bloodgood2013CoNLL, altschuler2019} with the regression approach in the current paper. 

\section{Conclusion \& Future Work} \label{sec:conclusion}

An area of interest in text classification is being able to forecast the performance of base learners in an iterative learning process. Past work has shown that forecasting models can be developed by regressing on a subset of the data that occurs before a cutoff we refer to as the $TPC$ and forecasting on the rest of the data. A critical question is what $TPC$ to use, which controls how early a forecast can be developed. In past work forecasts have been developed with a $TPC$ of fifteen percent of the data. We show in this paper that earlier forecasting would be beneficial. In many cases for text classification, forecasts can be developed with a $TPC$ between ten percent and fifteen percent of the data. However, analysis with active learning and stopping methods for active learning revealed that even earlier forecasting is still desired. We also found that forecasting is more difficult with some base learners than others, with decision tree classifiers being forecast the easiest, with SVM classifiers being in the middle, and neural network text classifiers being the hardest to forecast. We also found that using active learning algorithms made it harder to forecast due to the shape of the learning curve not matching current state of the art forecasting methods' expectations about the shape of learning curves. Finally, we found that forecasting performance is more difficult with some performance metrics than others. In particular, we found that forecasting performance measured by accuracy is much easier than forecasting performance measured by F-measure. Future work includes devising methods for even earlier forecasting of performance that are more accurate than the forecasting models currently used and better integrating forecasting methods with active learning stopping methods. 
\section*{Acknowledgment} \label{acknowledgment}

This work was supported in part by The College of New Jersey Support of Scholarly Activities (SOSA) program. The authors acknowledge use of the ELSA high performance computing cluster at The College of New Jersey for conducting the research reported in this paper. This cluster is funded by the National Science Foundation under grant number OAC-1828163.

\bibliographystyle{IEEEtran}
\bibliography{ms}

\end{document}